# Geologic context of the bright MARSIS reflectors in Ultimi Scopuli, South Polar Layered Deposits, Mars

**M.E. Landis[1]\* and J.L. Whitten[2]**

[1]Laboratory for Atmospheric and Space Physics, University of Colorado, Boulder, USA.

[2]Department of Earth and Environmental Sciences, Tulane University, New Orleans, USA.

Corresponding author: \*M.E. Landis (margaret.landis@lasp.colorado.edu)

**Key Points (140 characters or less):**

- No geologic unit nor surface feature matches with the location and orientation of the MARSIS bright reflector "lakes"
- Surface textures observed in HiRISE image data are relatively consistent across mapped geologic units
- Surface morphological mapping of the SPLD subglacial "lakes" region does not provide evidence consistent with the lake hypothesis

**Abstract (150 words or less)**

Radar-bright basal reflectors have been detected below the South Polar Layered Deposits (SPLD), Mars using Advanced Radar for Subsurface and Ionosphere Sounding (MARSIS) data and have an exciting but controversial interpretation: liquid water from subglacial lakes. We mapped the surface of the SPLD immediately above and surrounding the putative lakes (1:2M map scale) in order to provide geologic context for interpretation of the bright basal reflectors. We use THEMIS daytime IR (100 m/pixel), CTX (6 m/pixel), and HiRISE (25 cm/pixel) data to characterize geologic units and typical surface roughness. We find evidence for multiple geologic units with features due to $CO_2$ and aeolian-related processes. We do not find evidence for surface modification linked to the postulated lake locations. This is not consistent with the interpretation of the MARSIS basal radar reflector as subglacial lakes.

**Plain Language Summary**

Radar data are an important tool for understanding what lies below the surface of a geologic feature without having to sample it directly. Returned radar signal strength is higher typically due to a variety of factors, such as a strong difference in material (e.g., rock versus pure water ice) below the surface. A layer of material below the surface of Mars' south polar permanent ice sheet with unusually high returned radar signal has been detected from orbital data. One interpretation is that this is a subglacial lake, like those on Earth below the Antarctic ice sheet. Here, we use orbital data from two Mars spacecraft to make a map of the surface above the proposed subglacial lake. Our map shows many interesting features, but none that would have a clear subglacial lake origin.

## 1. Introduction

The South Polar Layered Deposits (SPLD) are a large water ice reservoir on Mars, composed of dusty ice similar to their northern counterpart. The SPLD have a surface exposure age that is orders of magnitude older than the NPLD's (e.g., Herkenhoff & Plaut, 2000; Koutnik et al., 2002). Adding to the complexity of the SPLD is the interpretation of bright Mars Advanced Radar for Subsurface and Ionosphere Sounding (MARSIS) basal radar reflectors in Ultimi Scopuli as a subglacial lake system (Lauro et al., 2021; Orosei et al., 2018). Hesperian-aged or older basal melting has been proposed for the south polar region (e.g., Ghatan & Head, 2004), but currently both PLD are interpreted as cold-based ice deposits (e.g., Fastook & Head, 2015; Fastook et al., 2012) and they do not show evidence of present-day flow (Koutnik et al., 2013; Winebrenner et al., 2008) and were unlikely to have flowed in the past (e.g., Smith & Holt, 2015). If present, subglacial lakes below the SPLD would rewrite our current understanding of the polar geology of Mars and the habitability of the planet.

Previous work has identified inconsistencies in the MARSIS subglacial lake hypothesis. The amount of salt required to either suppress the freezing point of water or for deliquescence to play the main role in maintaining this lake are non-physical (Sori & Bramson, 2019). While localized heating could generate liquid (Sori & Bramson, 2019), the highlands have inferred low mantle heat flow (Broquet et al., 2021; Ojha et al., 2021). The putative lake locations do not correspond to the minimum hydrological potential expected from basal topography (Arnold et al., 2019), potentially due to the relatively recent formation of the main lake (Orosei et al., 2018). However, detection of other lake locations (Lauro et al., 2021) makes the newly formed, still-adjusting lake system interpretation less likely due to the short timescales for melting to occur. The radar data may be as easily explained by more common martian materials, like hydrated clays or water ice, rather than liquid water (e.g., Bierson et al., 2021; Lalich et al., 2021; Smith et al., 2021). Overall, lake formation below the SPLD is challenging under our current understanding of martian geology and climate. Additional observational data are key to evaluating the hypothesis that there are extant lakes below the SPLD.

Subglacial lake detections in Antarctica rely on data from several different methods, including radar echo sounding (RES), ice surface topography, surface height changes, and seismic measurements (e.g., Wright & Siegert, 2011, and references therein). Terrestrial techniques used on the Greenland ice sheet (Jordan et al., 2018; Oswald et al., 2018) motivated Lauro et al. (2021) to investigate the radar reflection itself to understand the presence or absence of basal melt ponding. The presence of a subglacial lake may produce fracture or ridge features at the surface, like at Pine Island Glacier, Antarctica. This glacier shows fractures at the surface that have been explained by basal subglacial melt channels (Vaughan et al., 2012). The extent to which local fracture or subsidence features are present or absent would also provide a key observational constraint for future ice sheet stress modeling of these hypothesized lakes. Finally, in Greenland and Antarctica, ice cores are used to support the hypothesis that these unusual, remotely sensed reflectors result from melt. Ice cores are not currently available for Mars.

Understanding the geologic context of the bright MARSIS reflector (Orosei et al., 2018) is critical for its interpretation. The Hesperian-aged Dorsa Argentea Formation, surrounding and underneath the SPLD, corresponds to other bright MARSIS basal reflectors in many areas (Khuller & Plaut, 2021). The deposit may be a remnant of a previously more extensive ice sheet, where the ice content is now low compared to mid-latitude buried glaciers (Whitten et al., 2020). The SPLD shows an unusual and geographically extensive radar attenuation ("fog") and signal scattering at Shallow Radar (SHARAD) wavelengths when compared with the NPLD (Whitten

& Campbell, 2018). The SPLD have a larger average dust component than the NPLD (e.g., Plaut et al., 2007; Zuber et al., 2007), which could affect radar measurements. Further interpretation of SPLD radar scattering is hindered because the last extensive geologic mapping of the SPLD (Kolb & Tanaka, 2006) occurred before higher-resolution Mars Reconnaissance Orbiter (MRO) data were available.

Surface geology can be used to discern if landforms exist that indicate recent vertical movement of the ice, changes in basal stress conditions, or local melt areas potentially related to a subglacial lake. Higher resolution maps are required from data that has been collected from the High Resolution Imaging Science Experiment (HiRISE) (McEwen et al., 2007) (~25 cm/pixel) and continued daytime coverage from 2001 Mars Odyssey's Thermal Emission Imaging System (THEMIS) camera (Christensen et al., 2004) (~100 m/pixel) and MRO's Context Camera (Malin et al., 2007) (~6 m/pixel).

We map the surface geologic features and units above the lake region identified in Lauro et al. (2021) and Orosei et al. (2018) to answer three open questions that geologic mapping can uniquely address:

1. What is the geologic context of the putative "lakes"?

2. What are the morphologic features at the surface above the lake region? Are any consistent with major episodes of fracturing, subsidence, or compression/extension?

3. Do any observed morphologic features constrain when the last major disturbance to the surface occurred?

We present a regional geologic map (1:2M scale) with notional geologic units derived from THEMIS daytime IR mosaics, geomorphic descriptions using CTX image data, as well as roughness interpretations derived from HiRISE images to answer our questions. The detection of relatively abrupt changes in morphologic features (e.g., fractures, changes in surface texture) on the SPLD surface interior and exterior to the proposed lakes would serve as confirming evidence of a subglacial lake, and the apparent age of these features suggests when these lakes could have been active.

We do not find geologic units or surface features that uniquely coincide with the mapped SPLD subglacial lake. We do find a wide variety of surface alteration textures consistent with at least kyrs of surface modification. We do not find evidence from surface geology to clearly support a subglacial lake interpretation of the radar data, but cannot rule this hypothesis out. Our results provide constraints for any new geophysical modeling of the putative lakes.

## 2. Mapping

### 2.1. Regional Context

The previous geologic map of Kolb and Tanaka (2006) utilized THEMIS visible, Mars Orbiter Camera (MOC), and Mars Orbiter Laster Altimeter (MOLA) data to categorize Ultimi Scopuli as fully within the Planum Australe 2 (Aa2) unit. The Planum Australe 2 unit is <300 m and composed of partially deflated and pitted layers that are evenly bedded. Fully deflated sections of the Aa2 unit form a sublimation lag overlying the Planum Australe 1 (Aa1), a unit with thinner layers that are exposed at the surface in Promethei Lingula and in troughs near the $CO_2$-ice residual cap. The putative lake region near Ultimi Scopuli does not have any unique characteristics as defined by this geologic map, and therefore use of higher-resolution data is needed.

The region of interest (ROI) in Ultimi Scopuli (Orosei et al., 2018) overlies a spatially extensive low radar reflectivity zone (LRZ) (Phillips et al., 2011; Whitten & Campbell, 2018) that extends throughout the SPLD between 135º E and 255º E. This LRZ material is defined by SHARAD data and represents sections of the SPLD with few to no reflectors, indicating a homogenous layer of material. The LRZ in Ultima Lingula vary in thickness from 370 m at the western edge to ~80 m at their easternmost boundary. At the ROI, the LRZ are ~190 m thick, meaning that the <300 m thick Aa2 unit includes two defined radar facies (Whitten & Campbell, 2018): (1) LRZ and (2) an underlying layered unit known as the focused layer facies. The cause and compositional make-up of the non-residual cap LRZs is unknown (e.g., Phillips et al., 2011).

Generally, this ROI is dominated by landforms with different THEMIS IR pixel brightness, which is a measure of temperature due to illumination, thermophysical properties of the surface, and albedo. While there are scalloped areas (Grima et al., 2011) towards the northern half of the ROI, these are not elsewhere in the map region. These features have been proposed to have formed from low basal shear stress (Grima et al., 2011), which are conditions consistent with subglacial lake formation. Below the map scale, linear features occur at the 10s to 100s of meters scale. This general region has been identified as an area that hosts araneiforms (more colloquially, "spiders") (Schwamb et al., 2018).

## 2.2. Basemap data

We use the THEMIS day IR mosaic (~100 m/pixel), edition 12 available on the United States Geologic Survey (USGS) Astropedia repository (Edwards et al., 2011) to make our 1:2M scale geologic map of Ultimi Scopuli near the MARSIS bright basal reflector. This data set provides appropriate resolution for our 1:2M map scale. THEMIS IR data have been used extensively in geologic mapping efforts owing to their resolution and ability to show morphologic and thermophysical surface properties (e.g., Bleamaster & Crown, 2010; Mangold et al., 2004; Mest & Crown, 2014; Robbins & Hynek, 2012; Tanaka et al., 2014; Williams et al., 2009).

In this relatively flat region of the SPLD, changes in thermophysical properties are contributing more to the overall THEMIS day IR pixel brightness compared to illumination here than at other martian mapping sites. Given thermal properties of ice-cemented soil (Mellon et al., 2004) and water ice, the diurnal skin depths in SPLD-like material are ~20 cm and are ~5-6 m annually. Composition, including thermophysical properties, is a parameter that geologic maps of planetary surfaces can be expected to convey (e.g., Skinner et al., 2018). Therefore, the geologic units we identify reflect general thermophysical properties of the upper few decimeters to meters of the surface (Fig. 1) as well as morphological differences.

We used CTX (6 m/pixel) visible data after initial mapping with THEMIS (Fig. 1) to characterize the units further and to separate geomorphological versus thermophysical differences. We analyzed HiRISE data in the ROI to further examine surface textures on scales <100 m (Fig. S1). Follow-up HiRISE data were requested by the authors and acquired during MY 35 between solar longitude ($L_s$) ~270-340 to target the putative lake region and to cover our mapped THEMIS and CTX geologic units. A subset of these that included contact points between multiple geologic units and surface textures were selected for further study.

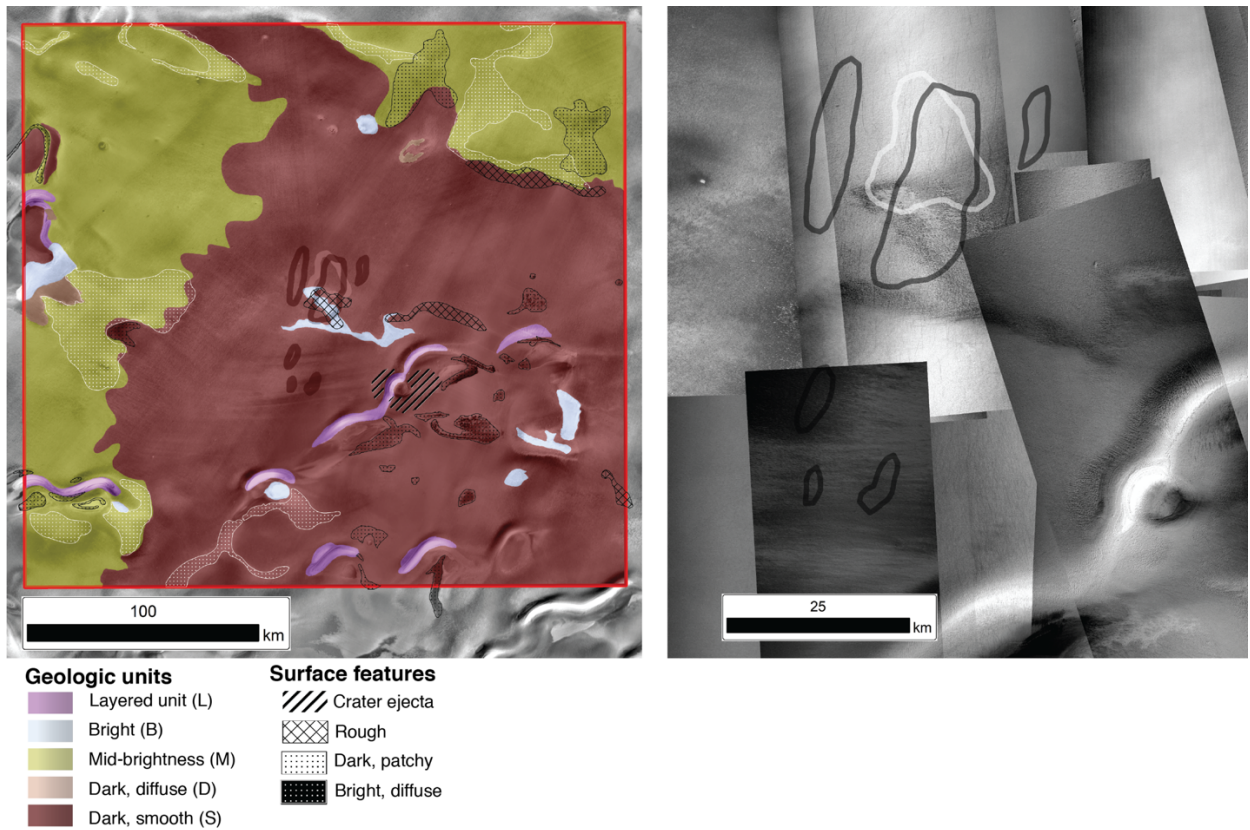

**Figure 1.** THEMIS basemap and geologic map (left) and example of CTX supplemental data from L$_s$ 230-10 (of the next MY) (right) over the subglacial lake region. Solid white lines represent the subglacial lake reported by (Orosei et al., 2018), solid black lines of the multiple subglacial lakes reported by (Lauro et al., 2021) (shown at 50% transparency). Both maps use a south pole stereographic projection for Mars (the south pole is to the up and right).

*2.3. Mapping methods*

The final map product has a scale of 1:2M (Supplemental Data Set S1). Mapping was

completed at the 1:500k scale (Skinner et al., 2018). The non-rectangular map area of ~240 km x

~260 km is defined the by corner coordinates: 212.0ºE, 82.1ºS; 178.7ºE, 83.3ºS; 201.2ºE, 78.4ºS;

179.1ºE, 79.2ºS (Fig. 1). Both authors mapped the region and compared their mapped units to generate a consensus version of the map in ESRI ArcMap 10.8. Once unit boundaries were identified in THEMIS data, we refined their descriptions using CTX visible data. The spatial location of unit boundaries identified in THEMIS are supported by CTX data.

## 2.4. Textural analysis at HiRISE scales

Surface feature identification was conducted at a map scale of 1:5k using HiRISE data (ESP_ 066074_0990, ESP_066602_0975, ESP_066786_0995, ESP_066866_0990). Spacing between landforms was measured at 1:15k map scale. All landforms mapped were troughs or topographic lows of varying morphology. Lines were drawn along the central part of trough floors to record their orientation and spacing. The spacing of the troughs was measured along lines that perpendicularly intersected the dominant orientation of the troughs. The bisecting lines were spaced ~700 to 800 m apart (Fig. S2).

## 3 Results and interpretation

### 3.1 Description of Geologic Units

We divide the surface into geologic units (large portions of the surface with a common description from the basemap and supplemental data) and surface features (which appear to modify or superpose the geologic units and are smaller in areal extent). When discussing bright/dark THEMIS IR data features, low/dark/colder is a pixel value of 1 and high/light/warmer is 255 in the grayscale product.

We identify five units (Fig. 1):

- Dark, diffuse margin unit (D): These are the darkest regions at THEMIS day IR wavelengths (12.57 microns). Their margins are poorly defined or diffuse. Individual deposits are relatively small, on the ~km scale. Material internal to the unit appears largely continuous, having a consistent pixel brightness in the THEMIS mosaic. These two factors distinguish unit D from the dark, patchy surface feature described below. These deposits have variable albedos in CTX. Dust devil tracks are apparent.

- Dark, smooth unit (S): This region is lighter at THEMIS day IR wavelengths than unit D. The margins vary from being sharp and well-defined to being more gradational. Generally, this unit is more continuous and extensive than unit D. In CTX data, the ridges are typically more muted than those in unit M, creating the appearance of a smooth surface at the map scale in both THEMIS and CTX data.

- Mid-brightness unit (M): This unit has an intermediate brightness between the darker units (D and S) and the brighter unit (B). The surface is smoother and more continuous than unit D, but more hummocky than unit S. Unit M occurs at the edges of the ROI and is interpreted to extend to the surrounding SPLD surface. In CTX data the surface is rough, with a high density of ridges.

- Bright unit (B): These regions of bright THEMIS day IR data appear in discrete units, either in roughly circular patches or in branched patches. The margins are generally well defined. In CTX, these deposits predominantly have either lower albedo contrast or have locally smooth surfaces.

- Layered unit (L): These areas, particularly near the walls of scalloped features, show horizontal bedding with brighter and darker layers. The CTX data are used to interpret this unit as exposures of multiple layers of the  SPLD in trough walls.

We identified four surface features based on THEMIS day IR brightness and their appearance as modifying or superposing one of the five main units:

1. Crater ejecta: These raised deposits form a detectable fan encircling a depression with a unique texture, interpreted as impact crater ejecta. This surface feature appears only once (171.90 ºE, 80.52 ºS) at the map scale.

2. Rough regions: Characterized by changes in roughness, usually with additional positive and negative topography, visible at the map scale in the THEMIS IR images. There is no accompanying brightness change in THEMIS day IR data that would indicate a change in background geologic unit. These surface features typically occur at the borders of geologic units.

3. Dark, patchy regions: These regions have a similar brightness to the S or D units. Quasi-circular or rounded features within these areas are common. Boundaries may be gradational, similar to the D unit, but the dark material in their interior appears patchy or discontinuous. Because surface variations are not visible in CTX, these features are likely in the near subsurface, and may be a thermophysical, rather than a morphologic signature.

4. Bright, diffuse regions: Compared with unit B, these regions appear slightly less bright and their margins are substantially more diffuse and gradational. The brightness of these surface features varies across each mapped location

## 3.2. Surface roughness

Surface roughness cannot be directly measured at scales relevant to MARSIS radar due to the limited availability of HiRISE digital elevation models in our ROI. Therefore, we map "roughness", or variability in surface textures, of our ROI using the spacing of surface features at the scale of HiRISE images. Four HiRISE images overlapping the map region were analyzed for morphology, orientation and spacing of small-scale landforms (Fig. 2) that include portions of the M, B, D, and S units (Fig. 1), spanning all units except for the L (layered) unit. All areas analyzed contain near-parallel troughs/depressions, though their morphologies vary between each HiRISE image. The morphology of the troughs immediately above the putative lakes is the most distinct (Fig. 2a), having jagged edges that we interpret to have been sculpted from secondary incision of the sides of the troughs. The walls are relatively steep, and the floors are relatively flat. Higher elevation material in between the troughs appears muted, with a few regions showing hints of polygonal patterned ground.

The remaining three regions contain troughs with smoother edges, creating a repeating parallel and slightly sinuous trough and ridge pattern. In ESP_066602_0975, the space between ridges has U-shaped floors covered with rough material, especially along the boundary between flat surface and ridge. The floor materials have a pattern that parallels the trough walls.

Polygonal patterned ground is also subtly visible (Fig. 2b). ESP_066786_0995 contains troughs with wider floors (Fig. 2c) compared with ESP_066602_0975. The entire region in this HiRISE image is covered with a polygonal pattern, with the polygons appearing larger on the ridge (or inter-trough) areas. Similar to ESP_066602_0975, ESP_066866_0990 has troughs with wide floors separated by ridges (Fig. 2d). The floors are covered with the largest observed polygons.

There are two dominant orientations for the mapped troughs, ENE and ESE (Fig. S4). Specifically, troughs are grouped into two orientations of 60º/240º and 110º/290º (north is 0º). These orientations are broadly consistent with predicted east-west wind directions in this portion of the SPLD (Howard, 2000). The ESP_066866_0990 and ESP_066602_0975 troughs are oriented ENE and both are on the west side of the map region. On the east side of the ROI, ESP_066786_0995 and ESP_066074_0990 have troughs with long axes oriented to the ESE. Spacing between troughs is consistent across the ROI, with an average spacing of 176±84 m (Fig. S5).

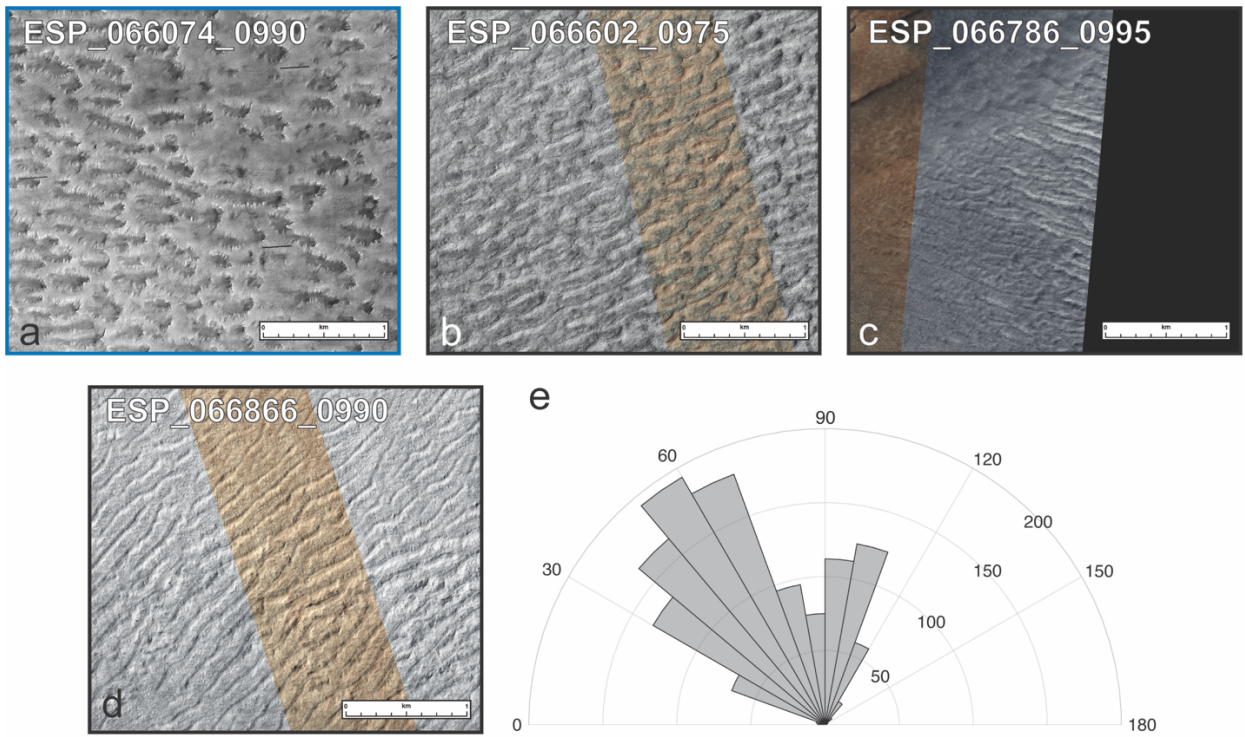

**Figure 2**. Close view of the trough features mapped in each HiRISE image and their orientations. (a) Blue border denotes that this region of troughs overlays the putative lakes. (e) Orientation of all measured troughs, where 0° is due north.

## 4 Discussion and interpretations

The purpose of mapping the Lauro et al. (2021); Orosei et al. (2018) region of interest was to answer three questions about the surface morphology and relative age of SPLD material above the putative subglacial lakes (section 1). We present the answers and our overall interpretations here.

### 4.1. What is the geologic context of the putative "lakes"?

The postulated outlines of the main and satellite lakes occur in a region close to the SPLD layers exposed in scarps (Layered or "L" unit), but are not near any other large change in topography at the map scale. The lakes are superposed by both the dark, smooth (S) and bright (B) units. The B unit, however, appears oriented perpendicularly (east-west orientation) to the direction of the lakes (south to north). Only the initial lake region as reported by Orosei et al. (2018) is below the B unit, and the remainder of the lakes underlie unit S, one of the most spatially extensive units in our mapping region. Unit B does have a change in surface roughness associated with it (see section 4.3), but this only intersects a portion of the total putative lake area. Therefore, while there are different geologic units that occur near the location of the lakes reported from MARSIS data and the B unit does appear at the location of the Orosei et al. (2018) putative lake, the orientation and size of the geologic units do not match these "lakes".

Ridge formations on the PLDs have a variety of proposed causes (Herny et al., 2014; Howard, 2000; Wilcoski & Hayne, 2019) that relate to wind and vapor conditions in the near-surface region. Changes in ridge pattern orientation and evolution is a current area of research, though these can largely be ruled out as having a subsidence or fracturing source at the 10s to 100 m scales.

*4.2 What are the morphologic features at the surface above the lake region? Are any consistent with major episodes of fracturing, subsidence, or compression/extension?*

Large linear features where a fracture or ridge in the surface appears at THEMIS scales do not occur in the ROI, including near the border of unit B or the lake region. Fractures around areas (that are not impact craters) where a sudden change in elevation occurred are also not present. Only troughs of varying morphologies were identified across the map region (Fig. 2).

Orosei et al. (2018) suggests that the SPLD ROI is too smooth to cause interference with MARSIS and therefore onboard processing was sufficient for the data. With our mapping of the periodic surface texture at HiRISE scales **(Fig. 2)**, we can reexamine that assumption. **Our work finds that the typical spacing between linear topographic features in the ROI varies from ~93-260 meters (Fig. S5). The wavelengths of MARSIS for subsurface sounding** in free space are ~60-160 m (Jordan et al., 2009), representing the vertical resolution of these instruments. Orosei et al. (2018) used the 3, 4 and 5 MHz MARSIS bands for bright basal reflector detection, meaning the relevant MARSIS wavelength range is ~60-100 m. The troughs would need to be spaced half a wavelength apart to create perfect constructive interference and substantially increase the echo strength of a radar reflector, e.g., ~30-50 m. The mapped troughs/depressions are too far apart to cause constructive interference.

*4.3 Do we observe signs of recent activity at the surface?*

In order to characterize signs of recent activity that could indicate active lake formation, we use observations of textures that have known or estimated formation times. Using impact craters is another possibility. However, the relatively small areal extent of geologic units (compared to the entire SPLD surface), combined with the small number and diameter of impact craters in the ROI means surface texture estimates provide the clearest relative age estimates.

The texture observed in unit B directly above the "lake" (Fig. 2a) is consistent with those formed by araneiforms, suggesting heavy local modification from solid state $CO_2$ ice processes through the solid-state greenhouse effect (Kieffer, 2007). Araneiforms have been previously observed near the map region of interest (Piqueux et al., 2003; Schwamb et al., 2018). Minimum formation times of araneiforms are ~ 1.3 kyr, though their periodic growth may mean they are much older (Portyankina et al., 2017). The presence of araneiforms suggests that significant local resurfacing driven by any subglacial effects could not have occurred in the last several thousand years.

Our ROI does overlap with a region identified as smooth on the kilometer-scale using MOLA data (Kreslavsky & Head, 2000). The MOLA-derived smooth area extends well beyond our ROI and is not unique to the putative lake region. If this region experienced major alteration due to a subglacial lake, the surface would have to have been modified significantly to re-smooth the region between the lake formation and present. The Myrs ages for the SPLD surface as a whole (e.g., Herkenhoff & Plaut, 2000) indicate that significant modification has not occurred. Therefore, if the subglacial lake does exist it must be modifying the surface more slowly than the development of araneiforms on the SPLD surface (e.g., Portyankina et al., 2017) or perhaps stopped Myrs ago.

Finally, the presence of ridges in some portions of the ROI provide constraints on the qualitative surface age. Wind-controlled availability of material to saltate or water vapor to diffuse may be an important factor on the SPLD. NPLD surface features have been suggested to form in a way related to snow sedimentation on Earth (e.g., megadunes, referred to as "sastrugi" in Herny et al., 2014). On the NPLD, linear surface textures may be explained by differences in surface illumination, and therefore temperature and accumulation due to vapor diffusion with time (e.g., Wilcoski & Hayne, 2020). The model of Wilcoski and Hayne (2020) suggests that ~10 m wavelength features form on kyr time scales. Our measured feature wavelengths of 100s of meters suggests a significantly longer time to develop these features. These are disrupted over the postulated "lakes" region and other locations in the ROI (e.g., Fig. 2c lower left corner). While not a strong constraint, it suggests that portions of the ROI are old enough to have developed araneiforms, and that any resurfacing significant enough to remove trough formation patterns has not occurred over kyr time scales. Therefore, knowing how surface textures form in icy environments where portions of the atmosphere freeze out as part of the seasonal cycle is key to further understanding this modification.

This region of the SPLD is also unique in katabatic wind patterns due to its relative lack of spiral troughs compared to the NPLD (Howard, 2000). Our ROI is dominated by E/W circumpolar winds, and the topographic variation of the large scallop just north of the ROI (Fig. 1) alters the wind direction (Howard, 2000). Dark linear features in CTX images are likely dust devil tracks (e.g., Bennett et al., 2017; Reiss et al., 2014), indicating aeolian processes may play an important role within the ROI and allowing for future potential comparisons between dust devil tracks and local prevailing wind conditions (e.g., Perrin et al., 2020).

Multiple lines of geologic evidence suggest that the surface of the ROI is at least kyr old, and that generally the SPLD surface records upwards of Myrs of geologic history. If subglacial melting had occurred and modified the surface, it must have done so more than a few kyr ago but any morphologic signs at the surface were unlikely to be totally erased in that time period if those features were larger than the ~meter-scale araneiforms.

*4.2 Interpretation*

The lack of substantial signs of recent disturbance at the surface in the mapped ROI is far from indisputable evidence of a lack of basal melting. However, without clear features that support any fracturing, subsidence, or change in basal stress state in the kyr to Myr past, it is worth examining alternative hypotheses for the unusual radar properties of the ROI.

The putative lake region is not the only region in the SPLD where there are bright basal reflectors (e.g., Khuller & Plaut, 2021; Plaut et al., 2007). Bright radar reflectors are identified at the base of the SPLD between approximately 305ºE and 0ºE, nearly coincident with the residual polar cap (e.g., Plaut et al., 2007), and are proposed to represent an interface between the SPLD and the Dorsa Argentea Formation (Plaut et al., 2007; Whitten et al., 2020). Exactly what materials are causing bright MARSIS radar reflectors across the SPLD remains an open question. MARSIS data processed in the same way as in Orosei et al. (2018) can produce bright reflectors over known equatorial volcanic terrains (Grima et al., 2022). Bright basal reflectors can be replicated by physical scenarios that require relatively ice-free material (e.g., Bierson et al., 2021; Lalich et al., 2021) or rely on ubiquitous materials on Mars, like smectite clays (Smith et al., 2021), though the exact radar properties of smectite clays are still being examined (Mattei et al., 2022). Research to distinguish between clay and brine basal compositions from radar is also underway on the Earth (e.g., Tulaczyk & Foley, 2020).

The discussion of alternative hypothesis to the lake interpretation of the bright MARSIS reflector begs the broader question of what makes the putative lake region different from the rest of the SPLD, if anything. The difference may be as simple as the SPLD overlying remnants of previous polar ice deposits composed of ice and various lithic components, like the North PLD and the cavi and rupes units (Tanaka et al., 2008), that may have various mixtures of ice and clays due to modification or mixing of materials in the past. The buildup and deflation of these ice sheets is evident in the NPLD (e.g., Smith et al., 2016; Tanaka et al., 2008) and by extension a similar, older history is likely recorded deep below the SPLD surface. Further exploration in radar analogs is needed to fully understand what produces localized bright basal radar reflectors, but liquid water is not required to replicate the MARSIS results, nor is it required to produce the geology at the surface of the SPLD.

## 5   Conclusions

Our mapped ROI, including the surface over the putative SPLD subglacial lakes, hosts a variety of surface textures and distinct geologic units. However, none of these geologic units show tell-tale signs of smoothing, compression/extension, or subsidence that uniquely correspond to the SPLD lakes. The surface morphology is likely the result of complex interactions between multiple processes that affect the surface evolution of the SPLD. More work is needed to understand the local and specific geologic histories of the near-surface region of the SPLD in general. Our geologic mapping of the region does not provide evidence to support the presence of a liquid water lake below the SPLD and places constraints on the timing and physical magnitude of surface features that could be plausibly generated by a subglacial lake.

## Acknowledgments and Data

Support for this work was provided by NASA PDART 80NSSC21K0882. The THEMIS daytime IR mosaic is available from the United States Geological Survey Astropedia (https://astrogeology.usgs.gov/search/map/Mars/Odyssey/THEMIS-IR-Mosaic-ASU/Mars_MO_THEMIS-IR-Day_mosaic_global_100m_v12). CTX and HiRISE image data are archived on the Planetary Data System (PDS) Imaging node (https://pds-imaging.jpl.nasa.gov/volumes/mro.html). Our geologic map and individual HiRISE images are included as a GIS layer package at [CU Scholar DOI TBD].

The authors thank Candice J. Hansen for helpful discussion on SPLD araneiforms, the HiRISE members for acquiring image data used in this study, Edgard Rivera-Valentín for useful discussion of salts and how to test the subglacial lake hypothesis, Shane Byrne and Kenneth E. Herkenhoff for helpful feedback on an early draft, and two anonymous reviewers during the peer-review process. We wish to acknowledge our peers in biological science research, healthcare and other essential workers whose hard work and sacrifice have allowed us to conduct this research during the COVID-19 pandemic.

**Geologic context of the bright MARSIS reflector in Ultimi Scopuli, South Polar Layered Deposits, Mars**

M.E. Landis[1] and J.L. Whitten[2]

[1]Laboratory for Atmospheric and Space Physics, University of Colorado, Boulder, USA.

[2]Department of Earth and Environmental Sciences, Tulane University, New Orleans, USA.

**Contents of this file**
        Figures S1 to S4

**Additional Supporting Information (Files uploaded separately)**

        Dataset S1, ESRI ArcGIS Layer Package for geologic map and HiRISE images used in this
        study

**Introduction**

Supporting information includes figures that note the locations of the postulated lakes and
new HiRISE data (Figure S1), as well as more detail on surface roughness calculations (Figure
S2 & S4) and orientations of particular features (Figure S3) presented in the main text.

Additional Supporting information includes Dataset S1, which contains a layer package for the
geologic map units as polygons and the HiRISE image data used in this study to further
characterize the surface of the region of interest. This layer package was generated using ESRI
ArcMap 10.8. The projection used is a south polar stereographic for Mars and projection
information is included in the HiRISE image data files.

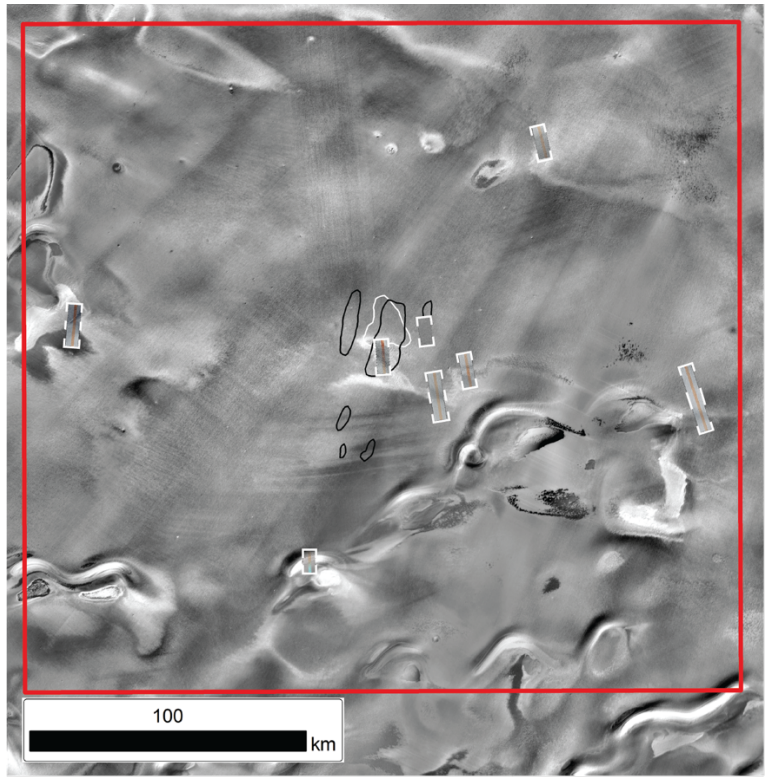

**Basemap:** Daytime THEMIS IR 100 m/pixel
**Supplemental data:** HiRISE images (l to r) ESP_066786_0995,
ESP_050566_1000, ESP_066074_0990, ESP_067644_0985,
ESP_067367_0990, ESP_067024_0990, ESP_066602_0975,
ESP_066866_0990

**Figure S1.** THEMIS basemap and supplemental HiRISE data footprints used to make the geologic map (Fig. 1) of the subglacial lake region. Solid white lines represent the subglacial lake reported by (Orosei et al., 2018), solid black lines of the multiple sub-glacial lakes reported by (Lauro et al., 2021), and the dashed white lines the locations of HiRISE images. This map uses a south pole stereographic projection for Mars (the south pole is to the up and right).

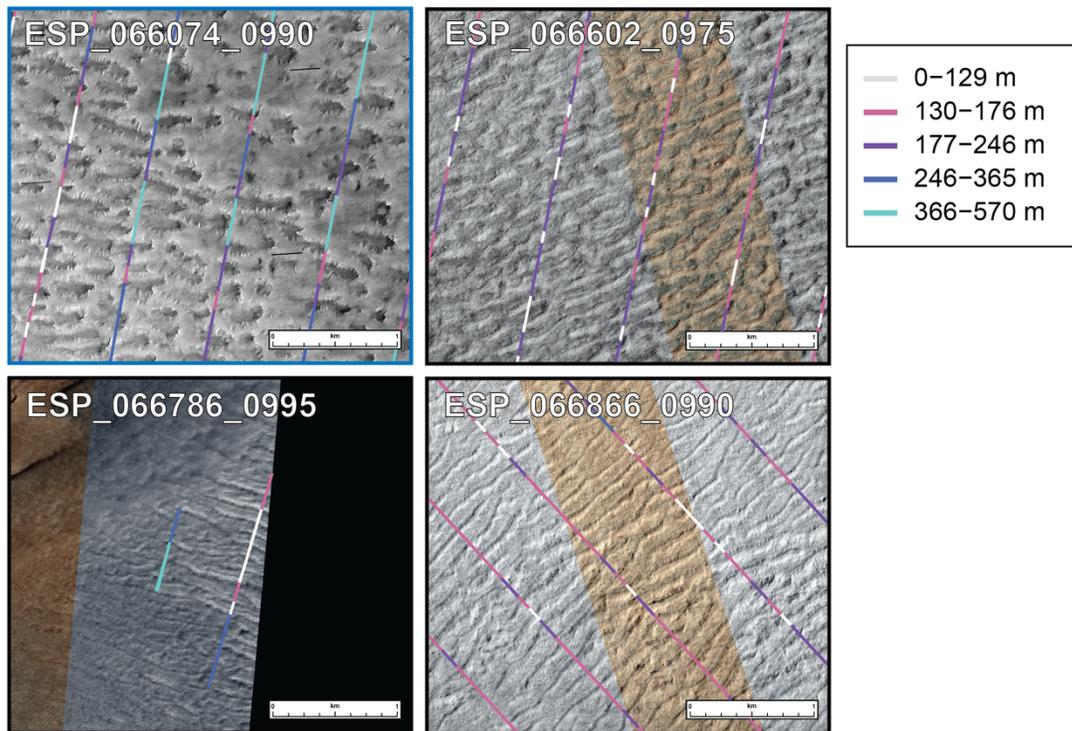

**Figure S2.** Example of how the spacing of troughs was determined. The changes in color in each line indicate the distance between troughs or topographic lows that intersect those lines. Parallel lines are spaced ~700-800 m apart (see main text, Section 2.4, for detailed description).

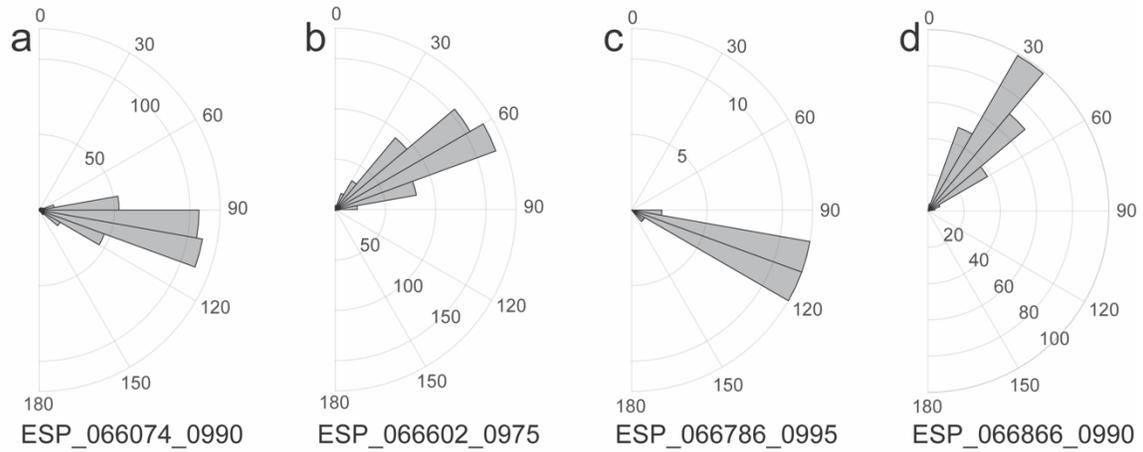

**Figure S3.** Orientation of the troughs in the greater putative lake region. (a) ESP_066074_0990 troughs, (b) ESP_066602_0975 troughs, (c) ESP_066786_0995 troughs, and (d) ESP_066866_0990 troughs above the proposed lake system.

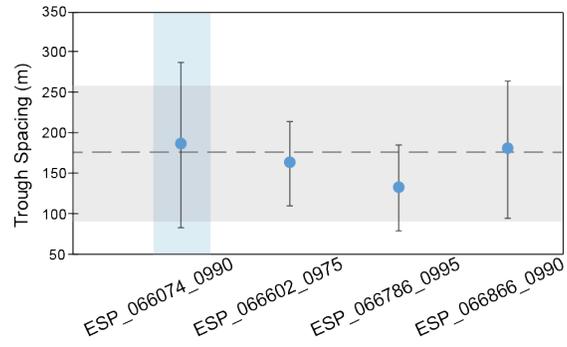

**Figure S4.** Average trough spacing in each of the four mapped HiRISE images. Gray dashed line and rectangle note the average spacing for all troughs and the associated 1σ value, respectively. The blue rectangle denotes the image overlying the putative lakes.